\documentclass{vienna-conf2019}
\usepackage{graphicx}
\usepackage{hyperref}
\usepackage[]{natbib}  
\usepackage{epstopdf}
\usepackage{amsmath,amssymb}

\def\BibTeX{{\rm B\kern-.05em{\sc i\kern-.025em b}\kern-.08em
    T\kern-.1667em\lower.7ex\hbox{E}\kern-.125emX}}
\bibpunct{(}{)}{;}{a}{}{,}  

\begin{document}

\TitreGlobal{Stars and their variability observed from space}


\title{Magnetic OB[A] Stars with TESS: probing their Evolutionary and Rotational properties - The MOBSTER Collaboration}

\runningtitle{MOBSTER}

\author{A. David-Uraz}\address{Department of Physics and Astronomy, University of Delaware, Newark, DE 19716, USA}

\author{C. Neiner}\address{LESIA, Paris Observatory, PSL University, CNRS, Sorbonne University, Université de Paris, 5 place Jules Janssen, 92195 Meudon, France}

\author{J. Sikora}\address{Physics and Astronomy Department, Bishop's University, Sherbrooke, QC J1M 1Z7, Canada}

\author{J. Barron$^{4,}$}\address{Department of Physics, Engineering Physics \& Astronomy, Queen's University, 64 Bader Lane, Kingston, ON K7L 3N6, Canada}\address{Department of Physics and Space Science, Royal Military College of Canada, PO Box 17000 Kingston, ON K7K 7B4, Canada}

\author{D.~M. Bowman}\address{Institute of Astronomy, KU Leuven, Celestijnenlaan 200D, B-3001 Leuven, Belgium}

\author{P. Cerraho\u{g}lu$^1$}

\author{D.~H. Cohen}\address{Department of Physics and Astronomy, Swarthmore College, Swarthmore, PA 19081, USA}

\author{C. Erba$^1$}

\author{V. Khalack}\address{D\'{e}partement de Physique et d'Astronomie, Universit\'{e} de Moncton, Moncton, NB E1A 3E9, Canada}

\author{O. Kobzar$^5$}

\author{O. Kochukhov}\address{Department of Physics and Astronomy, Uppsala University, Box 516, 75120, Uppsala, Sweden}

\author{H. Pablo}\address{American Association of Variable Star Observers (AAVSO), 49 Bay State Rd., Cambridge, MA 02138, USA}

\author{V. Petit$^1$}

\author{M.~E. Shultz$^1$}

\author{A. ud-Doula}\address{Penn State Scranton, 120 Ridge View Drive, Dunmore, PA 18512, USA}

\author{G.~A. Wade$^5$}

\author{the MOBSTER Collaboration}




\setcounter{page}{237}


\maketitle


\begin{abstract}
In this contribution, we present the MOBSTER Collaboration, a large community effort to leverage high-precision photometry from the Transiting Exoplanet Survey Satellite (\textit{TESS}) in order to characterize the variability of magnetic massive and intermediate-mass stars. These data can be used to probe the varying column density of magnetospheric plasma along the line of sight for OB stars, thus improving our understanding of the interaction between surface magnetic fields and massive star winds. They can also be used to map out the brightness inhomogeneities present on the surfaces of Ap/Bp stars, informing present models of atomic diffusion in their atmospheres. Finally, we review our current and ongoing studies, which lead to new insights on this topic.
\end{abstract}



\section{Introduction}

Magnetism is found in stars across the Hertzsprung-Russell diagram. However, while low-mass stars generate magnetic fields contemporaneously via strong surface dynamos powered by rotation and convection, stars across the OBA spectral type range lack the ingredients to actively create and sustain a large-scale magnetic field. Despite that fact, as evidenced by results from recent large spectropolarimetric surveys (e.g. MiMeS, BOB, BinaMiCS and LIFE; \citealt{2016MNRAS.456....2W, 2015IAUS..307..342M, 2015IAUS..307..330A, 2018MNRAS.475.1521M}), a small fraction ($\lesssim$ 10\%) of these stars exhibits the presence of strong, globally-organized magnetic fields on their surfaces. Even more remarkably, this incidence rate appears to be flat across a wide range of stellar masses, and the magnetic characteristics of these stars appear to be uncorrelated to their physical properties. The prevailing hypothesis to explain these observations is that the field is of \textit{fossil} origin, meaning that it was formed at an earlier stage of evolution \citep{1982ARA&A..20..191B}, although there remains a debate as to what that earlier stage might be \citep{2015IAUS..305...61N}.

The presence of a surface magnetic field influences several aspects of OBA stars that are crucial to their evolution. In the more massive O- and early B-type stars, magnetism can significantly spin down the star (e.g. \citealt{2009MNRAS.392.1022U}) and confine its wind, leading to much lower effective mass-loss rates \citep{2002ApJ...576..413U}. These effects greatly impact the evolution of magnetic OB stars and are starting to be included in evolutionary codes (e.g. \citealt{2019MNRAS.485.5843K}). The latter effect might even provide a viable channel to form heavy stellar-mass black holes (e.g. \citealt{2019arXiv191111989L}), such as those whose coalescence was detected by LIGO \citep{2016PhRvL.116f1102A}, at solar metallicity \citep{2017A&A...599L...5G, 2017MNRAS.466.1052P}.

In later B- and A-type stars, magnetic fields affect atomic diffusion processes in the atmosphere, leading to anisotropic chemical abundances on the stellar surface. This causes spectral peculiarities, as noted in the spectral type of many of these objects: the so-called ``Ap/Bp'' stars. When compared to other stars of similar spectral type, these are found to exhibit slower rotation rates as a population \citep{1995ApJS...99..135A}. In both cases, a further examination into the detailed interaction between magnetic fields and the atmospheres (and extended atmospheres) of these stars will better constrain their overall evolution and observable properties.

To this extent, we propose to take full advantage of the unique opportunity that the Transiting Exoplanet Survey Satellite (\textit{TESS}; \citealt{2015JATIS...1a4003R}) offers us. While it was originally designed, as its name suggests, for the detection and characterization of exoplanets via the transit method, the high-precision high-cadence data that it provides for objects covering $\sim$85\% of the sky represent a veritable treasure trove for the field of stellar astrophysics. In this article, we briefly describe what kind of photometric signature can be expected from magnetic massive and intermediate-mass stars, and what kind of physical insights we can expect to gain by analyzing \textit{TESS} data, and the ways in which we can increase that gain by combining them with various other observational diagnostics.

 



 
  
\section{General phenomenology}

Magnetic OBA stars tend to present a common photometric behavior. Their light curves are recognizable due to periodic rotational modulation. For the earlier spectral type stars, this is due to the magnetically confined wind material which scatters continuum light. Since this material is non-axisymmetrically distributed around the star and the magnetic and rotation axes are not generally aligned, a varying column density intersects the line of sight throughout a rotational cycle.

On the other hand, slightly cooler magnetic stars have, as mentioned above, abundance patches on their surfaces. Due to flux redistribution \citep{2007A&A...470.1089K}, these also lead to brightness inhomogeneities on the stellar surface, which appear and disappear as the star rotates, causing periodic photometric variations.

Both of these phenomena manifest themselves in a light curve periodogram as a well-defined peak in a reasonable frequency range that might be associated with the stellar rotation, and having at least one harmonic (typically the first one; \citealt{2018A&A...616A..77B}), although occultation by magnetospheric material can, in some cases, lead to the presence of more harmonics and a rather complex light curve \citep{2008MNRAS.389..559T}. This phenomenology can be used to select highly-probable magnetic candidates \citep{2018MNRAS.478.2777B}. However, this method must be used with caution since other mechanisms can lead to a similar observational signature (e.g. eclipsing binaries, ellipsoidal variations, etc.), and should be ruled out before the frequencies are ascribed to a rotational origin.

\section{Physical insights obtained from photometry}

\subsection{O and early B stars}

The circumstellar \textit{magnetospheres} formed by the interaction between a surface magnetic field and a strong stellar wind around OB stars have been studied in great detail over the past two decades. State-of-the-art magnetohydrodynamic simulations \citep{2002ApJ...576..413U, 2008MNRAS.385...97U, 2009MNRAS.392.1022U} have provided us with unprecedented insights into their structure and their influence on a star. Simpler parametrizations have also been developed, both for fast \citep{2005MNRAS.357..251T} and slow \citep{2016MNRAS.462.3830O} rotators, which allow us to define a density and velocity at any given point in the magnetosphere. Coupled with an appropriate radiative transfer scheme (e.g. \citealt{2018A&A...616A.140H}), these models can reproduce a large swath of multi-wavelength observations to great accuracy (e.g. \citealt{2014ApJS..215...10N, 2019MNRAS.483.2814D}). These can also be confronted with optical photometry, as was done with great success with MOST to explain the light curve of $\sigma$ Ori E \citep{2013ApJ...769...33T}. Ongoing efforts also aim to derive magnetospheric parameters using photometry \citep{2019MNRAS.tmp.2573M}. Finally, measuring changes in the periods of magnetic OB stars can validate or challenge our current understanding of their rotational evolution (e.g. \citealt{2010ApJ...714L.318T, 2011A&A...534L...5M, 2019MNRAS.486.5558S}).

\subsection{Late B and A stars}

Photometric time series can be used to map brightness spots on the surface of a star using various light curve inversion techniques, as exemplified by the work \citet{2016A&A...588A..54W} have done with BRITE data to map the surface of $\alpha$ Cir. These maps can then be compared to elemental abundance maps derived from Doppler Imaging applied to well-sampled time series of high-resolution spectroscopy (e.g. \citealt{1975Ap&SS..34..403K}). Furthermore, information about the abundance and vertical stratification of different chemical elements can be obtained by performing a detailed analysis of their line profiles, as is accomplished by the VeSElkA project \citep{2017MNRAS.471..926K}. There is a clear synergy between that endeavor and the core goals of the MOBSTER Collaboration leading to a more complete picture of the atmosphere of Ap/Bp stars, as evidenced by studies conducted in common by both groups \citep{2019MNRAS.490.2102K}.

\section{Progress to date}

So far, our collaboration has published three refereed articles, with more to come. Paper I \citep{2019MNRAS.487..304D} looked at the morphology of the photometric variations for known magnetic B- and A-type stars in sectors 1 and 2, concluding that most were compatible with rotational modulation. We also derived refined periods for those.

In Paper II \citep{2019MNRAS.487.4695S}, we detected rotational modulation in the light curves of many A-type stars observed by \textit{TESS} in sectors 1 to 4. Interestingly, the properties of these variations (e.g. period and amplitude) appear to differ between the population of stars identified as chemically peculiar (Ap) and the population of stars that are not identified as such. This might be providing us with a clue as to the underlying physical mechanisms.

Finally, in Paper III \citep{2019MNRAS.490.4154S}, we discuss the puzzling case of an eclipsing B-type binary system, HD 62658. Careful modelling of its light curve, as well as spectroscopic and spectropolarimetric data, suggests that only one of the stars is magnetic, even though the mass ratio is essentially unity and the stars appear to be otherwise nearly identical. This poses a particular challenge to theories of the origin of magnetic fields in these stars, thus thickening the plot even further.

\section{Conclusions and future work}

In conclusion, the MOBSTER Collaboration groups both observers and theorists in the goal of maximizing the scientific output of the \textit{TESS} mission with respect to magnetism in massive and intermediate mass stars. As elaborated above, the high-precision light curves produced by \textit{TESS} allow us to constrain the physics of the atmosphere and winds of these stars, leading to a deeper understanding of their evolution. While some work has already been done, there remain many questions to be answered. Furthermore, we have so far only skimmed the surface of this data bounty: we have focused mainly on 2-minute cadence data (which are obtained for somewhere between 200,000 and 400,000 objects), but full-frame image data (30 minute cadence) should be available for over 500 million point sources \citep{2018AJ....156..102S}. Thus, much work remains to unlock the truly transformative potential of the MOBSTER project in the field of magnetism in OBA stars.


\begin{acknowledgements}
ADU, VK and GAW acknowledge support from the Natural Sciences and Engineering Research Council of Canada (NSERC). Some of the research leading to these results has received funding from the European Research Council (ERC) under the European Unions Horizon 2020 research and innovation programme (grant agreement No. 670519: MAMSIE). CE acknowledges graduate assistant salary support from the Bartol Research Institute in the Department of Physics, University of Delaware, as well as support from program HST-GO-13629.002-A that was provided by NASA through a grant from the Space Telescope Science Institute. MES acknowledges financial support from the Annie Jump Cannon Fellowship, endowed by the Mount Cuba Observatory and supported by the University of Delaware. AuD acknowledges support from NASA through Chandra Award number
TM7-18001X issued by the Chandra X-ray Observatory Center, which is
operated by the Smithsonian Astrophysical Observatory for and on behalf
of NASA under contract NAS8-03060.
\end{acknowledgements}

\bibliographystyle{aa}  
\bibliography{david-uraz_9o04} 

\end{document}